\documentclass[11pt,a4paper]{article}

\usepackage{amsmath,amssymb}
\usepackage{bm}

\def\Pexp{\mathop{\rm Pexp}\nolimits}
\def\tr{\mathop{\rm tr}\nolimits}

\newcommand{\RR}{\mathbb{R}}
\newcommand{\ol}{\overline}

\begin{document}

\begin{titlepage}
\title{
\vspace{-1.5cm}
\begin{flushright}
{\normalsize TIT/HEP-699\\ March 2024}
\end{flushright}
\vspace{1.5cm}
\LARGE{Giant graviton expansions for line operator index}}
\author{
Yosuke {\scshape Imamura\footnote{E-mail: imamura@phys.titech.ac.jp}},
\\
\\
{\itshape Department of Physics, Tokyo Institute of Technology}, \\ {\itshape Tokyo 152-8551, Japan}}

\date{}
\maketitle
\thispagestyle{empty}
\begin{abstract}
We discuss giant graviton expansions for the Schur index
of ${\cal N}=4$ $U(N)$ SYM with the insertion of Wilson lines of the fundamental and
the anti-fundamental representations.
We first propose a double-sum giant graviton expansion
and numerically confirm that it correctly reproduces the line-operator index.
We also find that it reduces to a simple-sum expansion when we treat
the index as a Taylor series with respect to a specific fugacity.
\end{abstract}

\end{titlepage}

\tableofcontents

\section{Introduction}

The superconformal index
\cite{Romelsberger:2005eg,Kinney:2005ej}
of ${\cal N}=4$ $U(N)$ SYM is defined by
\begin{align}
\tr\left[(-1)^Fq^{J_1}p^{J_2}x^{R_x}y^{R_y}z^{R_z}\right]
\label{sci}
\end{align}
where $J_1$ and $J_2$ are spins and $R_x$, $R_y$, and $R_z$ are $R$-charges.
To preserve supersymmetry we impose the constraint $qp=xyz$ on the fugacities.
This can be regarded as the partition function in the spacetime $S^3\times S^1$.

In this paper we are interested in the index with the insertion of Wilson line operators
\cite{Dimofte:2011py,Gang:2012yr,Drukker:2015spa,Hatsuda:2023iwi,Guo:2023mkn,Hatsuda:2023imp,Hatsuda:2023iof}.
We insert $1/2$-BPS Wilson lines of fundamental and anti-fundamental representations
at the north and south poles of $S^3$, respectively.
This breaks $so(4)$ rotational symmetry of $S^3$ to $so(3)$.
We define the poles so that they are fixed under $J_1$ action.
Then, $J_1$ is the Cartan generator of the preserved rotational symmetry,
which we denote by $so(3)_{J_1}$.
The insertion also breaks $so(6)_R$ symmetry to $so(5)_R$,
and without loosing generality we assume that
$R_x$ and $R_y$ are preserved.
The symmetries generated by $J_2$ and $R_z$ are broken,
and we have to turn off the corresponding fugacities $p$ and $z$.
Then the superconformal index reduces to the Schur index \cite{Gadde:2011uv}.
In the following we use $I_N$ for the Schur index without line operator insertion
\begin{align}
I_N=\tr\left[(-1)^Fq^{J_1}x^{R_x}y^{R_y}\right],\quad
q=xy.
\end{align}
See \cite{Bourdier:2015wda,Pan:2021mrw,Hatsuda:2022xdv} for analytic expressions of $I_N$ with finite $N$.

We denote the Schur index with the Wilson line operator insertion by $I_{{\rm line},N}$.
$I_{{\rm line},N}$ can be calculated by the
localization formula
\begin{align}
I_{{\rm line},N}=\int_{U(N)}d\mu \left[\Pexp\left\{\left(1-\frac{(1-x)(1-y)}{1-q}\right)\chi_{\rm adj}\right\}
\chi_{\rm fund}\ol\chi_{\rm fund}
\right].
\label{iline}
\end{align}
$\chi_{\rm adj}$,
$\chi_{\rm fund}$, and
$\ol\chi_{\rm fund}$
are the $U(N)$ characters of the adjoint, the fundamental, and the anti-fundamental representations, respectively.
$\Pexp$ is the plethystic exponential and $\int_{U(N)}d\mu$ is the
integral over gauge group $U(N)$ with the Haar measure.
The factor $\chi_{\rm fund}\ol\chi_{\rm fund}$ in the integrand corresponds to the line operator insertion,
and $I_N$ can be calculated by (\ref{iline}) with the factor removed.
We can also consider Wilson line operators associated with more general representations
and the corresponding index can be calculated by replacing $\chi_{\rm fund}\ol\chi_{\rm fund}$
in (\ref{iline})
by the
product of the corresponding characters.
In this paper, however, we will not discuss such a generalization.

It is well known that in the large $N$ limit the superconformal index
$I_\infty$ (even before taking the Schur limit)
coincides with the index of the supergravity multiplet in $AdS_5\times S^5$ \cite{Kinney:2005ej}
via AdS/CFT correspondence \cite{Maldacena:1997re,Gubser:1998bc,Witten:1998qj}.
It is in the Schur limit given by
\begin{align}
I_\infty=I_{\rm sugra}=\Pexp i_{\rm sugra}
\end{align}
with the supergravity letter index
\begin{align}
i_{\rm sugra}
=\frac{x}{1-x}
+\frac{y}{1-y}
-\frac{q}{1-q}.
\end{align}

The large $N$ limit of the line-operator index was calculated in \cite{Gang:2012yr}.
See also \cite{Hatsuda:2023imp,Hatsuda:2023iof}.
On the AdS side, the line operators correspond to
a fundamental string extended along $AdS_2\subset AdS_5$ \cite{Rey:1998ik,Maldacena:1998im}.
Two line operators are identified with the two boundaries of the
worldsheet.
$I_{{\rm line},\infty}$ is reproduced
by taking account of the contribution from
fluctuation modes on the string worldsheet \cite{Gang:2012yr}:
\begin{align}
I_{{\rm line},\infty}
=I_{\rm sugra}I_{\rm F1}=\Pexp(i_{\rm sugra}+i_{\rm F1}),
\label{ilineholo}
\end{align}
where $i_{\rm F1}$ is the letter index of worldsheet fluctuations
\begin{align}
i_{\rm F1}
=x+y-q.
\end{align}
The explicit form of the index is
\begin{align}
I_{{\rm line},\infty}
&=1+4t+10t^2+22t^3+44t^4+82t^5+146t^6+\cdots,
\label{ilineinf}
\end{align}
where we used the unrefinement $x=y=t$ to simplify the equation.
The variable $t$ is also used for order counting.
${\cal O}(t^n)$ indicates terms of the form $x^{n-k}y^k$ or higher power terms.

The purpose of this paper is to generalize the holographic relation (\ref{ilineholo})
to finite $N$
by including contributions of 
giant gravitons \cite{McGreevy:2000cw,Grisaru:2000zn,Hashimoto:2000zp,Mikhailov:2000ya}.

\section{Leading corrections}
Let us first look at some numerical results for small $N$:
\begin{align}
I_{{\rm line},0}&=0,\nonumber\\
I_{{\rm line},1}&=1+2t+t^2+2t^3+2t^4+0t^5+3t^6+\cdots,\nonumber\\
I_{{\rm line},2}&=1+4t+7t^2+8t^3+11t^4+16t^5+15t^6+\cdots,\nonumber\\
I_{{\rm line},3}&=1+4t+10t^2+18t^3+25t^4+36t^5+52t^6+\cdots,\nonumber\\
I_{{\rm line},4}&=1+4t+10t^2+22t^3+39t^4+58t^5+87t^6+\cdots,\nonumber\\
I_{{\rm line},5}&=1+4t+10t^2+22t^3+44t^4+76t^5+117t^6+\cdots,\nonumber\\
I_{{\rm line},6}&=1+4t+10t^2+22t^3+44t^4+82t^5+139t^6+\cdots.
\end{align}
Again we showed unrefined indices.
We defined $I_{{\rm line},0}=0$ because we cannot insert
non-trivial line operators in $N=0$ theory.
(In other words, we cannot introduce strings ending on D-branes
if there are no D-branes.)
These are converging to (\ref{ilineinf}) in the large $N$ limit,
and the leading term of the finite $N$ correction is
\begin{align}
I_{{\rm line},N}-I_{{\rm line},\infty}
&=-(N+1)t^N+{\cal O}(t^{N+1}).
\label{lineleading}
\end{align}
This is similar to the finite $N$ correction of $I_N$ \cite{Bourdier:2015wda}:
\begin{align}
I_N-I_\infty=-(N+2)t^{N+1}+{\cal O}(t^{N+2}).
\label{schurleading}
\end{align}
Let us first review how
(\ref{schurleading}) is reproduced on the AdS side \cite{Arai:2019xmp,Arai:2020qaj}.
In \cite{Arai:2020qaj} the following expansion for the Schur index was proposed.
\begin{align}
\frac{I_N}{I_\infty}=\sum_{m_x,m_y=0}^\infty
x^{m_xN}F^{(x)}_{m_x}
\cdot
(xy)^{m_xm_y}
\cdot
y^{m_yN}F^{(y)}_{m_y}.
\label{schurgge}
\end{align}
The non-negative integers $m_x$ and $m_y$ are
wrapping numbers of giant gravitons, and this kind of
expansions are named in \cite{Gaiotto:2021xce}
giant graviton expansions.
See also \cite{Imamura:2021ytr,Murthy:2022ien,Lee:2022vig,Beccaria:2023zjw,Beccaria:2024szi}.

The summand in (\ref{schurgge}) consists of three factors.
The first factor
$x^{m_xN}F^{(x)}_{m_x}$ is the
contribution from
a stack of $m_x$ D3 giants wrapped around the $R_x$-fixed locus in $S^5$.
$x^{m_xN}$ comes from the $R$-charge of the giants,
and $F^{(x)}_{m_x}$ is the Schur index of the field theory realized on the
worldvolume of the giants.
The last factor $y^{m_yN}F^{(y)}_{m_y}$ is a similar contribution from
$m_y$ giants on the $R_y$-fixed locus in $S^5$.
Because the worldvolumes of the giants are $S^3\times S^1$,
which are isomorphic to the AdS boundary,
$F^{(x)}_m$ and $F^{(y)}_m$ are essentially the same as $I_m$.
Due to the different actions of the symmetry generators
they are related by \cite{Arai:2020qaj}
\begin{align}
F^{(x)}_m=\sigma_{x}I_m,\quad
F^{(y)}_m=\sigma_{y}I_m,
\label{ffromi}
\end{align}
where $\sigma_x$ and $\sigma_y$ are the following simple variable changes.
\begin{align}
\sigma_x:(x,y)\rightarrow(x^{-1},xy),\quad
\sigma_y:(x,y)\rightarrow(xy,y^{-1}).
\label{sxsy}
\end{align}
The explicit forms of $F^{(x)}_N$ for small $N$ are \cite{Arai:2020qaj}
\begin{align}
F^{(x)}_0&=1,\nonumber\\
F^{(x)}_1
&=\frac{-x}{1-\frac{y}{x}}+x^2(-1+\tfrac{y}{x})+x^3(-1+\tfrac{y^3}{x^3})
+x^4(-1+\tfrac{y^2}{x^2}-\tfrac{y^3}{x^3}+\tfrac{y^4}{x^4})+\cdots
\nonumber\\
F^{(x)}_2
&=\frac{\frac{x^5}{y}(-1+2\frac{y}{x})}{(1-\frac{y}{x})(1-\frac{y^2}{x^2})}
+\tfrac{x^7}{y^2}(-1+2\tfrac{y^2}{x^2})
+\tfrac{x^9}{y^3}(
-1
+2\tfrac{y^3}{x^3}
+2\tfrac{y^6}{x^6}
)
+\cdots
\nonumber\\
F^{(x)}_3
&=\frac{\frac{x^{12}}{y^3}(-2+3\frac{y}{x}+3\frac{y^2}{x^2}-5\frac{y^3}{x^3})}{(1-\frac{y}{x})(1-\frac{y^2}{x^2})(1-\frac{y^3}{x^3})}
+\frac{\frac{x^{15}}{y^5}(-2+3\frac{y^2}{x^2}+3\frac{y^3}{x^3}-5\frac{y^5}{x^5})}{1-\frac{y^2}{x^2}}+\cdots.
\end{align}
The same expansions were also obtained in \cite{Beccaria:2024szi}
using analytic results in \cite{Hatsuda:2023iwi}.

The middle factor
in (\ref{schurgge}),
$(xy)^{m_xm_y}$, is the contribution from the bi-fundamental
fields arising from open strings stretched between the two loci.
Although the letter index of the bi-fundamental fields depends on the gauge fugacities,
its plethystic exponential becomes the factor independent of
the gauge fugacities.
Due to this accidental simplification, the gauge fugacity integrals of $U(m_x)$ and $U(m_y)$
gauge groups are factorized.

Let us focus on the leading correction from $(m_x,m_y)=(1,0)$ and $(0,1)$.
\begin{align}
\frac{I_N}{I_\infty}=1+x^NF^{(x)}_1+y^NF^{(y)}_1+\cdots,
\end{align}
where $F^{(x)}_1$ is the index of the $U(1)$ vector multiplet
on a giant graviton wrapped around the $R_x$-fixed locus:
\begin{align}
F^{(x)}_1
&=\sigma_xI_1
=\Pexp\frac{\frac{1}{x}-2y+xy}{1-y}
\nonumber\\
&=\Pexp(\tfrac{1}{x}+\tfrac{y}{x}+{\cal O}(t^1)).
\label{funcf}
\end{align}
The first term $\frac{1}{x}$ and the second term $\frac{y}{x}$ in the letter index
are
${\cal O}(t^{-1})$ and ${\cal O}(t^0)$, respectively,
and these two terms are enough to obtain the leading term in the $t$-expansion of $F^{(x)}_1$
\begin{align}
F^{(x)}_1=\frac{1}{1-\frac{1}{x}}\frac{1}{1-\frac{y}{x}}(1+{\cal O}(t^1))=\frac{-x}{1-\frac{y}{x}}+{\cal O}(t^2).
\end{align}
The negative order term $\frac{1}{x}$ in the letter index corresponds to the unwrapping mode (tachyonic mode)
of the giant graviton.
Its plethystic exponential $\frac{1}{1-\frac{1}{x}}=\frac{-x}{1-x}$ includes the factor $-x$,
which has two effects: it changes the overall sign and it raises the order of the leading correction
by one.
This is the mechanism generating the negative leading correction of ${\cal O}(t^{N+1})$ in
(\ref{schurleading}).

Let us turn to the leading correction of $I_{{\rm line},N}$ shown in (\ref{lineleading}).
A common feature with (\ref{lineleading}) is that it is negative.
We expect this is again due to the unwrapping modes because even when we
introduce the string worldsheet corresponding to the line operators
giant gravitons still have unwrapping modes.
However, the order of (\ref{lineleading}) seems contrary to the fact that the unwrapping modes raises the order
by one.
To match the order of the correction, we need extra factor of ${\cal O}(t^{-1})$.
One possible source of this factor is a backreaction of the introduction of the
string worldsheet.
Unfortunately, we cannot give a clear explanation for this factor at present,
and in the following, we simply assume the existence of such a factor.

Now, let us start the analysis of giant graviton contributions to $I_{{\rm line},N}$.
We consider the situation in which the string worldsheet and a single giant graviton
coexist.
There are two cases.
One is the case in which the string and the giant graviton do not correlate.
Namely, the string does not end on the worldvolume of the giant.
The other is the case that the string worldsheet is divided into two
parts by the giant graviton and the two semi-infinite strings are ending on the worldvolume of the giant.
In the former case, we obtain the contribution
\begin{align}
I_{{\rm line},N}=\cdots+I_{\rm sugra}(x^NF_1^{(x)}+y^NF_1^{(y)})I_{\rm F1}+\cdots.
\end{align}
We also have similar contributions from other wrapping numbers $(m_x,m_y)$,
and they are summed up to $I_NI_{\rm F1}$.
Let us subtract this contribution from $I_{{\rm line},N}$.
Furthermore, we remove the supergravity contribution by dividing the
difference by $I_{\rm sugra}=I_\infty$.
Numerical analysis on the gauge theory side shows
\begin{align}
\frac{I_{{\rm line},N}-I_NI_{\rm F1}}{I_\infty}=x^NF'^{(x)}+y^NF'^{(y)}+{\cal O}(t^{2N+2}),
\label{fpxfpy}
\end{align}
where $F'^{(x)}$ and $F'^{(y)}$ are $N$-independent functions.
$F'^{(x)}$ is
\begin{align}
F'^{(x)}
&=\frac{-1}{1-\frac{y}{x}}+x\frac{-1-\frac{y^2}{x^2}}{1-\frac{y}{x}}+x^2\frac{-1+\frac{y}{x}+\frac{y^2}{x^2}-\frac{y^3}{x^3}-\frac{y^4}{x^4}}{1-\frac{y}{x}}
\nonumber\\
&+x^3(-1+2\tfrac{y^3}{x^3}+2\tfrac{y^4}{x^4}+\tfrac{y^5}{x^5})
+x^4(-1-\tfrac{y^3}{x^3}+3\tfrac{y^5}{x^5}+2\tfrac{x^6}{y^6}+\tfrac{x^7}{y^7})
\nonumber\\
&+{\cal O}(t^5)
\label{fprime}
\end{align}
and $F'^{(y)}$ is obtained from $F'^{(x)}$ by exchanging $x$ and $y$.
We can rewrite (\ref{fprime}) in the simple form
\begin{align}
F'^{(x)}
&=\frac{1}{x}\Pexp\left(\frac{\frac{1}{x}-2y+xy}{1-y}+2(y-xy)\right).
\label{fprime2}
\end{align}
This is close to (\ref{funcf}).
The first term in the parentheses
is the same as the letter index in (\ref{funcf}),
and interpreted as the contribution from
a giant graviton wrapped on the $R_x$-fixed locus.
The prefactor $1/x$ is the mismatch factor mentioned above,
and we have no clear explanation for it.
We accept the existence of such factors,
and later we will insert
such factors by hand.

The remaining part, $2(y-xy)$ in the parentheses
is expected to be the contribution from
the two semi-infinite string worldsheets ending on the
giant graviton.
In the next section we will show that this
is indeed obtained from the mode analysis on
the semi-infinite strings.

\section{Worldsheet fluctuations}
Fluctuation modes on a superstring worldsheet extended along $AdS_2\subset AdS_5$ are
studied in \cite{Drukker:2000ep,Faraggi:2011bb}.
They form a short multiplet of $osp(4^*|4)$, and the spectrum is
\begin{align}
\varphi(1,\bm{1},\bm{5})
+\psi(\tfrac{3}{2},\bm{2},\bm{4})
+\phi(2,\bm{3},\bm{1}).
\label{ultrashort}
\end{align}
$\varphi$ and $\phi$ are
scalar fields corresponding to five directions in $S^5$
and three transverse directions in $AdS_5$, respectively.
$\psi$ is the fermion field with $8$ degrees of freedom.
We showed
in (\ref{ultrashort})
the representations under $so(2,1)_{\rm conf}\times so(3)_{J_1}\times so(5)_R$.
The value $\nu$ in the first slot in each representation is
the energy eigenvalue of the conformal primary state normalized by the AdS radius.
In addition, there are infinitely many conformal descendants with
energies $\nu+1,\nu+2,\ldots$ in each conformal representation.

The presence of a giant graviton wrapped around the $R_x$-fixed locus breaks
$so(5)_R$ to $so(2)_{R_x}\times so(3)_{R_y}$.
Under this symmetry breaking $\varphi$ and $\psi$
split into $(\varphi_3,\varphi_X,\varphi_{\ol X})$ and $(\psi_+,\psi_-)$, respectively.
See Table \ref{rncomponents} for quantum numbers of each component.

The $so(2,1)_{\rm conf}\times so(3)_{J_1}\times so(5)_R$ quantum numbers of the supercharges
preserved by the string worldsheet are $(\pm\tfrac{1}{2},\bm{2},\bm{4})$.
Among the real $16$ supercharges only eight with $H=R_x$ are preserved after the introduction of a giant
wrapped around the $R_x$-fixed locus,
and the superconformal representation (\ref{ultrashort})
is decomposed into small representations labeled by $n=H-R_x=0,1,2,\ldots$.
We denote them by ${\cal R}_n$, and their components are shown in Table \ref{rncomponents}.
\begin{table}[htb]
\caption{Components of the representation ${\cal R}_n$ and their quantum numbers are shown.
The last column shows the mode functions of the scalar fields.}\label{rncomponents}
\centering
\begin{tabular}{ccccccc}
\hline
\hline
&$E$ & $so(3)_{J_1}$ & $R_x$ & $so(3)_{R_y}$ & $n$ & mode \\
\hline
$\varphi_{\ol X}$ & $n-1$ & $\bm{1}$ & $-1$ & $\bm{1}$ & $n\geq2$ & $f_{n-2}^{(1)}$ \\
$\psi_-$ & $n-\frac{1}{2}$ & $\bm{2}$ & $-\frac{1}{2}$ & $\bm{2}$ & $n\geq2$ \\
$\phi$ & $n$ & $\bm{3}$ & $0$ & $\bm{1}$ & $n\geq2$ & $f^{(2)}_{n-2}$ \\
$\varphi_3$ & $n$ & $\bm{1}$ & $0$ & $\bm{3}$ & $n\geq1$ & $f^{(1)}_{n-1}$ \\
$\psi_+$ & $n+\frac{1}{2}$ & $\bm{2}$ & $+\frac{1}{2}$ & $\bm{2}$ & $n\geq1$ \\
$\varphi_{\ol X}$ & $n+1$ & $\bm{1}$ & $+1$ & $\bm{1}$ & $n\geq0$ & $f^{(1)}_n$ \\
\hline
\end{tabular}
\end{table}

Before the introduction of the giant graviton,
there are three BPS modes
saturating the bound $H-J_1-R_x-R_y\geq0$
and contributing to the index.
One of them 
with $(H,J_1,R_x,R_y)=(1,0,+1,0)$ belongs to ${\cal R}_0$,
and the other two with
$(H,J_1,R_x,R_y)
=(1,0,0,+1)$ and
$(\tfrac{3}{2},+\tfrac{1}{2},+\tfrac{1}{2},+\tfrac{1}{2})$
belong to ${\cal R}_1$.
These three modes give the three terms $x$, $y$, and $-xy$
in $i_{\rm F1}$, respectively.

When the string worldsheet is attached on
the worldvolume of the giant graviton,
some representations ${\cal R}_n$ are removed by
the boundary conditions.
To determine which of them are allowed,
let us explicitly write down the mode functions for
the scalar fields.

The scalar fields satisfy the wave equations
\begin{align}
(\Delta_{AdS_2}-2)\phi=0,\quad
\Delta_{AdS_2}\varphi=0.
\label{kgeqs}
\end{align}
These equations can be easily solved as follows.
We define the $AdS_2$ as a subspace in $\RR^{2,1}$
with the metric
\begin{align}
ds^2=-2dzdz^*+dx^2
\end{align}
by the embedding
\begin{align}
z=\tfrac{1}{\sqrt{2}}\cosh\rho e^{it},\quad
z^*=\tfrac{1}{\sqrt{2}}\cosh\rho e^{-it},\quad
x=\sinh\rho.
\end{align}
We introduce conformal generators
\begin{align}
K&=i(x\partial_{z^*}+z\partial_x),\nonumber\\
P&=i(x\partial_{z}+z^*\partial_x),\nonumber\\
H&=-z\partial_z+z^*\partial_{z^*}.
\end{align}
satisfying 
\begin{align}
[H,P]=P,\quad
[H,K]=-K,\quad
[K,P]=H.
\end{align}
The mode function $f_0^{(\nu)}$ of the conformal primary state
satisfying $Kf_0^{(\nu)}=0$ and $Hf_0^{(\nu)}=\nu f_0^{(\nu)}$ is given by
\begin{align}
f_0^{(\nu)}(t,\rho)=\frac{1}{z^\nu}\propto\frac{e^{-i\nu t}}{\cosh^\nu\rho}.
\end{align}
The mode functions $f_k^{(\nu)}$ for conformal descendant states are obtained by applying an arbitrary number of
the raising operator $P$ to $f_0^{(\nu)}$;
\begin{align}
f_k^{(\nu)}(t,\rho)=P^kf_0^{(\nu)}(t,\rho)\quad
k=0,1,2,\ldots.
\end{align}
Using the relation between
the scalar field Laplacian $\Delta_{AdS_2}$
and the Casimir operator of the conformal algebra
\begin{align}
\Delta_{AdS_2}
=H^2-(PK+KP),
\end{align}
we can easily find that the mode functions $f^{(\nu)}_k$ satisfy
\begin{align}
\Delta_{AdS_2}f_k^{(\nu)}=\nu(\nu-1)f_k^{(\nu)},
\end{align}
and $\phi$ and $\varphi$ satisfying
(\ref{kgeqs}) can be expanded by $f_k^{(2)}$ and $f_k^{(1)}$,
respectively.

An important fact is that the mode functions
$f^{(\nu)}_k$ are even for even $k$ and
odd for odd $k$ under the reflection of the worldsheet $\rho\rightarrow-\rho$.
This means that for a field with Dirichlet (Neumann) boundary condition
only modes with odd (even) $k$ are allowed.
We can easily find that scalar modes in ${\cal R}_n$ with odd $n$
satisfy the boundary conditions imposed on the giant graviton
worldvolume, and the fluctuations in
${\cal R}_n$ with even $n$ are not allowed.

Concerning the three BPS saturating modes,
only the two in ${\cal R}_1$ are consistent with the boundary
conditions, and they give the letter index $y-xy$.
Furthermore, we have two semi-infinite strings ending on the
giant.
Therefore, total contribution to the index is
\begin{align}
I^{(x)}_{\rm F1}=\Pexp i^{(x)}_{\rm F1},\quad
i^{(x)}_{\rm F1}=2(y-xy).
\label{msemi}
\end{align}
With the presence of the factor $1/x$,
this correctly reproduces (\ref{fprime2}).
\begin{align}
F'^{(x)}=\frac{1}{x}F^{(x)}_1I^{(x)}_{\rm F1}.
\end{align}
$F'^{(y)}$ is also obtained in a similar way.

\section{Multiple-wrapping contributions}
Now, based on the analysis of the single-wrapping contribution
in the previous sections,
we propose the following giant graviton expansion
for $I_{{\rm line},N}$.
\begin{align}
\frac{I_{{\rm line},N}
}{I_\infty}
&=
I_{\rm F1}
\sum_{m_x,m_y=0}^\infty
x^{m_xN}F^{(x)}_{m_x}
\cdot
(xy)^{m_xm_y}
\cdot
y^{m_yN}F^{(y)}_{m_y}
\nonumber\\
&+\sum_{m_x,m_y=0}^\infty
x^{m_xN}F^{(x)}_{{\rm line},m_x}\frac{I_{\rm F1}^{(x)}}{x}
\cdot
(xy)^{m_xm_y}
\cdot
y^{m_yN}F^{(y)}_{m_y}
\nonumber\\
&+\sum_{m_x,m_y=0}^\infty
x^{m_xN}F^{(x)}_{m_x}
\cdot
(xy)^{m_xm_y}
\cdot
y^{m_yN}F^{(y)}_{{\rm line},m_y}\frac{I_{\rm F1}^{(y)}}{y}.
\label{ggeline}
\end{align}
For each pair of wrapping numbers
$(m_x,m_y)$ we have to consider three contributions.
The first line comes from
giants on which string worldsheet does not end.
The second line is the contribution
from giants with semi-infinite strings ending on the $R_x$-fixed locus,
and the third line comes from giants with semi-infinite strings ending on
the $R_y$-fixed locus.
We do not have to consider the case with two semi-infinite
strings ending on different loci because
the contribution from such a configuration
vanishes when we perform the factorized gauge fugacity integrals.

$\frac{I_{\rm F1}^{(x)}}{x}$ in the second line is the contribution from
semi-infinite strings
(\ref{msemi}) together with the mismatch factor $1/x$.
$F^{(x)}_{{\rm line},m_x}$ is the
contribution from the field theory realized on the worldvolume of
$m_x$ coincident giants on the $R_x$-fixed locus.
We take account of the attached strings as the
line operator insertion.
This is not necessary when $m_x=1$ because
the effect of the insertion is trivial
and $F^{(x)}_{{\rm line},1}=F^{(x)}_1$.
We define $F^{(x)}_{{\rm line},0}=F^{(y)}_{{\rm line},0}=0$.
For $(m_x,m_y)=(1,0)$ the second line
in (\ref{ggeline})
reduces to the single-wrapping contribution
$x^NF'^{(x)}$ in (\ref{fpxfpy}).

$F^{(x)}_{{\rm line},m}$ and $F^{(y)}_{{\rm line},m}$ are related to
$I_{{\rm line},m}$ by the variable changes
(\ref{sxsy}).
\begin{align}
F^{(x)}_{{\rm line},m}=\sigma_xI_{{\rm line},m},\quad
F^{(y)}_{{\rm line},m}=\sigma_yI_{{\rm line},m}.
\label{flinefromiline}
\end{align}
It is not so easy to perform these variable changes
because we need analytic continuation \cite{Gaiotto:2021xce,Imamura:2022aua}.
Here we use the method in \cite{Arai:2020qaj}.
Namely, we go back to the
localization formula
(\ref{iline})
with the variable changes (\ref{sxsy}) applied to the integrand, and
perform the integral with
contours carefully chosen in the consistent way with
the analyticity.
The results for small $N$ are
\begin{align}
F^{(x)}_{{\rm line},0}
&=0,\nonumber\\
F^{(x)}_{{\rm line},1}
&=F^{(x)}_1,
\nonumber\\
F^{(x)}_{{\rm line},2}
&=\frac{-\frac{x^4}{y}}{1-\frac{y^2}{x^2}}
+\frac{\frac{x^6}{y^2}(-1-\frac{y}{x}+5\frac{y^2}{x^2}-\frac{y^3}{x^3}-\frac{y^4}{x^4}+\frac{y^5}{x^5})}{(1-\frac{y}{x})(1-\frac{y^2}{x^2})}
\nonumber\\&
+\frac{\frac{x^8}{y^3}(-1-2\frac{y}{x}+6\frac{y^3}{x^3}+\frac{y^4}{x^4}-4\frac{y^5}{x^5}-\frac{y^8}{x^8})}
{1-\frac{y^2}{x^2}}
\nonumber\\&
+\tfrac{x^{10}}{y^4}(-1-2\tfrac{y}{x}-\tfrac{y^2}{x^2}+5\tfrac{y^4}{x^5}-4\tfrac{y^7}{x^7}+\tfrac{y^8}{x^8}+\tfrac{y^{10}}{x^{10}})
+{\cal O}(t^7),\nonumber\\
F^{(x)}_{{\rm line},3}
&=\frac{-\frac{x^{10}}{y^3}}{1-\frac{y^3}{x^3}}
+\frac{\frac{x^{13}}{y^5}(-1-2\frac{y^2}{x^2}+\frac{y^3}{x^3}+5\frac{y^4}{x^4}-\frac{y^6}{x^6}+\frac{y^8}{x^8})}{(1-\frac{y^2}{x^2})(1-\frac{y^3}{x^3})}
+{\cal O}(t^9).
\label{fline0123}
\end{align}
$F^{(y)}_{{\rm line},N}$ are obtained from these by exchanging $x$ and $y$.
It is demanding to calculate higher order terms in $F_{{\rm line},3}^{(x)}$,
and we have obtained only the first two terms shown in (\ref{fline0123}).
(The method in \cite{Beccaria:2024szi} may be useful to compute these functions more efficiently.)
With this result we checked (\ref{ggeline})
correctly reproduces the index up to ${\cal O}(t^{3N+7})$ for
$N\leq3$.

\section{Simple-sum expansion}
As was first found in \cite{Gaiotto:2021xce},
the double-sum expansion (\ref{schurgge}) for the Schur index
and the triple-sum expansion in 
\cite{Imamura:2021ytr}
for the superconformal index
reduce to simple-sum expansions
associated with only one cycle
if we adopt an appropriate expansion scheme.
See also \cite{Imamura:2022aua,Fujiwara:2023bdc} for the mechanism
of the reduction to simple-sum expansions.
It is the case for the line-operator index, too.
If we treat the indices as $y$-series rather than $t$-series,
\begin{align}
I_N(x,y)=\sum_{k=0}^\infty f_{N,k}(x)y^k,\quad
I_{{\rm line},N}(x,y)=\sum_{k=0}^\infty f_{{\rm line},N,k}(x)y^k,
\label{yseries}
\end{align}
then the contributions from $m_y\geq 1$ decouple,
and the double-sum expansion (\ref{ggeline})
reduces to the simple-sum expansion
\begin{align}
\frac{I_{{\rm line},N}}{I_\infty}
&=
I_{\rm F1}
\sum_{m_x=0}^\infty
x^{m_xN}F^{(x)}_{m_x}
+
\frac{I_{\rm F1}^{(x)}}{x}
\sum_{m_x=0}^\infty
x^{m_xN}F^{(x)}_{{\rm line},m_x}
\label{simplesum}
\end{align}
associated with only the $R_x$-fixed locus.
The variable change $\sigma_x$ in (\ref{sxsy})
acts independently on each term in the $y$-series (\ref{yseries}),
and
replace the coefficient functions $f_{N,k}(x)$ and $f_{{\rm line},N,k}(x)$
by $x^k f_{N,k}(x^{-1})$ and $x^k f_{{\rm line},N,k}(x^{-1})$, respectively.
Because the coefficient functions are rational functions,
it is easy to obtain $F_N^{(x)}$ and $F_{{\rm line},N}^{(x)}$ from $I_N$ and $I_{{\rm line},N}$ by
the relations in (\ref{ffromi}) and (\ref{flinefromiline}).
In particular, the leading terms in
(\ref{yseries}), which gives the Coulomb branch limit $y\rightarrow 0$ of the indices,
are given by
\begin{align}
f_{N,0}=\prod_{i=1}^N\frac{1}{1-x^i},\quad
f_{{\rm line},N,0}
=\frac{1}{1-x}\prod_{i=1}^{N-1}\frac{1}{1-x^i},
\end{align}
and it is easy to analytically prove
the relation (\ref{simplesum}) in the Coulomb branch limit.
One can also numerically check that
(\ref{simplesum}) holds even without taking the Coulomb branch limit.

\section{Discussion}
In this paper we discussed the Schur index with the
insertion of fundamental and anti-fundamental line operators
for ${\cal N}=4$ $U(N)$ SYM with finite $N$.
As expected, we found that the finite $N$ corrections
are reproduced by summing up the contributions from
giant gravitons.
Unfortunately, we have not yet have clear explanation for the
factors $1/x$ and $1/y$ appearing when
the semi-infinite strings
end on giant gravitons.
A potential origin of the factors is a back-reaction of the
fundamental strings.

An important and interesting problem is
the generalization to other representations of the line operators.
It is known that symmetric and anti-symmetric representations
with rank $k$ of order $N$ are holographically realized by
a D3-brane wrapped on $AdS_2\times S^2$ \cite{Drukker:2005kx} and
a D5-brane wrapped on $AdS_2\times S^4$ \cite{Yamaguchi:2006tq}, respectively.
More general representations with the rank $k$ of order $N^2$
are realized by bubbling geometries \cite{DHoker:2007mci}.
In the large $N$ limit
the matching of the indices for these cases were
studied in \cite{Gang:2012yr,Hatsuda:2023iof}.
It is interesting to study how the finite $N$ corrections
arise on the AdS side as giant graviton contributions.

Another important direction is to consider line operator indices in other theories \cite{Gang:2012yr,Guo:2023mkn}:
${\cal N}=4$ SYM with $SO$ and $Sp$ gauge groups and ${\cal N}=2$ SCFTs which have holographic duals.
Giant graviton expansions for such theories without operator insertions have been studied in
\cite{Arai:2019xmp,Arai:2019wgv,Fujiwara:2023bdc},
and we expect they can be generalized to line operator indices.

In this paper we proposed the giant graviton expansions
based on the analysis on the AdS side.
It is also important to understand how the expansion arises
on the gauge theory side 
in a similar way to the derivation of the giant graviton expansion
for the superconformal index in \cite{Gaiotto:2021xce}.

We look forward to revisiting these aspects in the near future.

\section*{Acknowledgments}
The author thanks Tadashi Okazaki and Daisuke Yokoyama
for valuable discussions.
This work was supported by JSPS KAKENHI Grant Number JP21K03569.

\end{document}